# Cryptocurrency bubbles, the wealth effect, and non-fungible token prices: Evidence from metaverse LAND

Kanis Saengchote*

*Chulalongkorn University*

This version: 9 September 2022

***ABSTRACT***

The rapid rise of cryptocurrency prices led to concerns (e.g. the Financial Stability Board) that this wealth accumulation could detrimentally spill over into other parts of the economy, but evidence is limited. We exploit the tendency for metaverses to issue their own cryptocurrencies along with non-fungible tokens (NFTs) representing virtual real estate ownership (LAND) to provide evidence of the wealth effect. Cryptocurrency prices and their corresponding real estate prices are highly correlated (more than 0.96), and cryptocurrency prices Granger cause LAND prices. This metaverse bubble reminisces the 1920s American real estate bubble that preceded the 1929 stock market crash.

* Chulalongkorn Business School, Chulalongkorn University, Phayathai Road, Pathumwan, Bangkok 10330, Thailand. (email: kanis@cbs.chula.ac.th).

0

# 1. Introduction

Crypto Investors Are Wealthier. No One Knows How Much They're Spending.

It's time to start contemplating how vast wealth created in cryptocurrencies filters through the rest of the economy.

*Telis Demos, Wall Street Journal, February 18, 2022.*

The rapid rise in cryptocurrency and crypto asset prices that began in late 2020 led to significant wealth accumulation that could spill over into other parts of the economy. CoinGecko, a crypto asset information aggregator, reported that at the beginning of July 2020, total crypto market capitalization stood at $262.8 million. By mid-November 2021, the figure rose to more than $3 trillion. From some, this is a boon: news articles in December 2021 often discussed how this newfound wealth is related to the boom in luxury goods spending.[1] But for others, this could be a bane: a report by the Financial Stability Board (FSB) released in February 2022 explicitly outlined the "wealth effect" as a vulnerability, where "changes in the value of crypto-assets might impact their investors, with subsequent knock-on effects in the financial system". As the market collapsed in May 2022, the narrative changed from glamour stories of nouveau riche to how luxury watches previously in shortage were now flooding the market again.[2]

While the wealth effect is a potential systemic risk concern, during the same period, other risky asset classes such as equity also experienced similar movements. In an IMF blogpost in early 2022, Adrian, Iyer and Qureshi (2022) commented that "there's a growing interconnectedness between virtual assets and financial markets" and showed that the correlation between Bitcoin and the S&P 500 index increased from almost zero in 2017 to 0.34 in 2020 – 2021.[3] This co-movement makes it difficult to disentangle the crypto wealth effect from a general wealth effect. In addition, a February 2022 Wall Steet Journal article suggested that not only it is hard to observe how the crypto millionaires are spending, but such wealth is also hard to spend because crypto assets remain largely unconnected to the financial system, making it challenging to establish the wealth effect spillover.

However, for crypto assets recorded directly onto blockchains, the exchange and settlement processes are simpler because they exist in the same system, making it easier to spend the accumulated wealth by acquiring on-chain assets. The rise of crypto market capitalization coincides with an interest in non-fungible tokens (NFTs), unique on-chain units of digital storage that could be used to hold a variety of information including artwork and proof of virtual real estate ownership. Thus, the co-movements between related crypto assets can provide better identification of the wealth effect.

---

[1] https://www.bloomberg.com/news/articles/2021-12-15/cryptocurrency-helping-spur-u-s-as-luxury-s-new-growth-driver, accessed on September 9, 2022.
[2] https://www.bloomberg.com/news/articles/2022-07-29/the-crypto-collapse-has-flooded-the-market-with-rolex-and-patek, accessed on September 9, 2022.
[3] https://blogs.imf.org/2022/01/11/crypto-prices-move-more-in-sync-with-stocks-posing-new-risks/, accessed on September 9, 2022. Note: the IMF uses the term "virtual assets" to describe transferable digital information recorded on blockchains. Other common terms used are "digital assets" and "crypto assets".

0

In a related study, Dowling (2022b) analyzed the relationship between cryptocurrencies (unbacked, blockchain-based assets, e.g. Bitcoin and Ether) and three types of NFTs (virtual real estate, digital art and game characters) and found strong evidence of co-movement. We build upon this finding by exploiting a direct relationship between NFTs and cryptocurrencies in virtual real estate platforms (also referred to as "metaverse"), as they tend to issue unbacked, freely floating crypto assets intended as medium of exchange in their corresponding metaverse. They also issue NFT proof of ownership, often referred to as "LAND". For example, Decentraland – one of the earliest and most successful metaverses developed on the Ethereum blockchain – issues MANA as its cryptocurrency, while The Sandbox – also on Ethereum – issues SAND.[4] Buyers can potentially use LAND to conduct side transactions include leases, but as of August 2022, there is no way to directly earn real estate income in both metaverses.

In this article, we analyze the returns spillover between cryptocurrencies and LAND NFTs to directly investigate the wealth effect, which is also often the focus for research on the economics of art (Goetzmann, Renneboog and Spaenjers, 2011; Pénasse and Renneboog, 2022). Recent articles have documented the economics and returns properties of metaverse real estate (e.g. Goldberg et al., 2021; Dowling, 2022a; Nakavachara and Saengchote, 2022) and their relationship with cryptocurrencies (Dowling, 2022b), but linkages are often analyzed from the perspective of volatility spillovers (e.g. Bouri, Lucey and Roubaud, 2020; Moratis, 2021; Dowling, 2022b), while our focus is on the co-movements in prices and returns.

Our analysis is divided into two steps: first, we document metaverse cryptocurrency price bubbles that could lead to accumulation of wealth using a timestamping algorithm developed by Phillips, Shi and Yu (2015). Researchers have found that cryptocurrencies are prone to media attention (Philippas et al., 2019), psychological fear of missing out (Baur and Dimpfl, 2018) and herding behavior (Bouri, Gupta and Roubaud, 2019), thus they are prone to bubbles. Then, we use the vector autoregression (VAR) model and Granger causality test to examine the lead-lag relationship as evidence of directional wealth spillover effect. Our article is also related to a growing literature on detecting and explaining asset bubbles, which have been conducted in cryptocurrencies (e.g. Corbet, Lucey and Yarovaya, 2018; Bouri, Shahzad and Roubaud, 2019; Shahzad, Anas, and Bouri, 2022), real estate (DeFusco, Nathanson and Zwick, 2022), equity (Phillips, Shi and Yu, 2015; Liao, Peng and Zhu, 2022) and art (Pénasse and Renneboog, 2022).

## 2.    Data and Methodology

Daily cryptocurrency price data is obtained from CoinGecko. Prices quoted in USD are used in the bubble timestamping analysis and as exchange rates to convert the price of LAND NFT transactions which commonly occur in non-USD denomination. Summary statistics of cryptocurrency prices are reported in Table 1.

---

[4] Details of Decentraland can be found in Goldberg et al. (2021), while details of The Sandbox can be found in Nakavachara and Saengchote (2022).

**Table 1: Daily Cryptocurrency Prices**
This table presents the summary statistics of daily cryptocurrency prices obtained from CoinGecko between January 1, 2021 and August 12, 2022. CoinGecko obtains trading data from various exchanges around the world and reports the volume-weighted average price calculated using data screened by its proprietary outlier detection algorithm. MANA is the cryptocurrency of Decentraland and SAND is the cryptocurrency of The Sandbox, both of which are built on the Ethereum blockchain. Bitcoin (BTC) is the cryptocurrency native to the Bitcoin blockchain, while Ether (ETH) is the cryptocurrency native to the Ethereum blockchain. Percentage of days bubble-stamped are dates identified by the Phillips, Shi and Yu (2015) algorithm as explosive bubbles.

|                    | MANA | SAND | BTC    | ETH   |
|--------------------|------|------|--------|-------|
| Mean               | 1.45 | 1.71 | 42,389 | 2,622 |
| Min                | 0.08 | 0.04 | 19,047 | 730   |
| Median             | 0.98 | 0.86 | 42,202 | 2,604 |
| Max                | 5.20 | 7.51 | 67,617 | 4,815 |
| Standard deviation | 1.10 | 1.72 | 11,481 | 962   |
| Skewness           | 1.04 | 1.28 | -0.05  | 0.22  |
| Kurtosis           | 3.19 | 3.70 | 2.36   | 2.16  |
| % Bubble-stamped   | 5.6% | 5.3% | 0.0%   | 3.0%  |

We obtain secondary LAND transactions of Decentraland and The Sandbox directly from Ethereum.[5] Transactions occurred from August 2018 for Decentraland and December 2019 for The Sandbox to August 2022. While blockchains allow users to directly interact with one another, the inherent anonymity makes peer-to-peer exchanges challenging. Decentraland maintains its own marketplace, while The Sandbox relies on Internet-based third-party providers, where OpenSea is the dominant marketplace: 21% and 96% of transactions on Decentraland and The Sandbox during the sample period occur on OpenSea. Each recorded transaction on the blockchain may involve transfers of multiple of LAND plots for a single payment. Consequently, we count the number of plots involved in the transaction and include it as a control variable. Transaction prices are winsorized at thresholds of 0.1% and 99.9% to limit the influence of outliers. There are 17,118 and 47,385 unique transactions in Decentraland and The Sandbox respectively.

Transactions paid in wETH – a "wrapped" version of ETH – are flagged as they represent a different type of transaction. OpenSea allows both direct purchases (fixed-price NFTs) and bidding on auctions (offers). Ether (ETH) is Ethereum's native coin created via the consensus algorithm that secures the integrity of the blockchain similar to Bitcoin (BTC), and wETH is a digital "token" created by smart contract program and issued by depositing ETH in exchange for wETH. While ETH and wETH are 1:1 pegged, the nature of the data is not the same. Sellers on OpenSea have choices regarding which cryptocurrency to list their NFTs in, but buyers must use wETH to "make an offer on an ETH item" because of this technical difference.[6] Consequently, transactions conducted in wETH (rather than ETH) are more likely to be from auction bids rather than fixed-price purchases, and thus may have different prices. Table 2 summarizes LAND transactions used in the study: 7.3% of Decentraland's transactions are settled in wETH, while the proportion is higher at 20.1% for The Sandbox, reflecting the greater reliance of The Sandbox on Opensea.

---
[5] The Ethereum addresses for the LAND contracts are '0xf87e31492faf9a91b02ee0deaad50d51d56d5d4d' for Decentraland and '0x50f5474724e0ee42d9a4e711ccfb275809fd6d4a' (the legacy contract created in December 2019) and '0x5cc5b05a8a13e3fbdb0bb9fccd98d38e50f90c38' (the new address with upgraded functionality created in January 2022) for The Sandbox.
[6] https://support.opensea.io/hc/en-us/articles/360063518053-How-do-I-make-an-offer-on-NFTs-

**Table 2: LAND Transactions**
This table reports the summary statistics of secondary LAND transactions of Decentraland and The Sandbox obtained from the Ethereum blockchain. Transactions occurred from August 2018 for Decentraland and December 2019 for The Sandbox to August 2022. Prices are converted to USD using daily prices obtained from CoinGecko. Number of plots is the number of LAND NFTs that change ownership in a single transaction. Transactions settled in wETH (smart contract, "wrapped" version of Ether) are more likely to correspond to auction bids rather than fixed-price purchases and thus are explicitly flagged.

|  | Decentraland | | The Sandbox | |
| --- | --- | --- | --- | --- |
| Sample size | 17,118 | | 47,385 | |
| Paid in wETH | 7.30% | | 20.10% | |
|  | USD price | Num plot | USD price | Num plot |
| Mean | 7,790.67 | 1.06 | 9,140.90 | 1.64 |
| StdDev | 19,083.38 | 0.76 | 15,790.93 | 2.50 |
| Skewness | 15.70 | 18.77 | 5.92 | 5.51 |
| Kurtosis | 322.81 | 433.77 | 46.17 | 43.51 |
| p5 | 450.71 | 1 | 139.03 | 1 |
| p50 | 3,471.74 | 1 | 5,789.95 | 1 |
| p95 | 20,779.96 | 1 | 21,021.45 | 9 |

## 2.1 Bubble Timestamping

To identify phases of bubble, we use the Phillips, Shi and Yu (2015) (PSY) bubble timestamping algorithm which conducts a series of augmented Dickey and Fuller (1979) (ADF) unit root test on a time series. The algorithm calculates the ADF test statistic from the regression:

$$\Delta y_t = \alpha_{s_1,s_2} + \beta_{s_1,s_2} y_{t-1} + \sum_{i=1}^{k} \delta_{s_1,s_2}^k \Delta y_{t-1} + \varepsilon_t \qquad (1)$$

Where $y_t$ is the time series of interest, $k$ is the number of lags, and $s_1$ and $s_2$ are the starting and ending points used for the estimation (thus, $s_2 - s_1$ is the window width). Let $ADF_{s_1,s_2}$ denote the associated test statistic. For simplicity, we can normalize the sample period to be between interval $[0,1]$. Let $r_0$ denote the last observation of the time series used in the first regression (hence, the minimum width), then the backward supremum ADF (BSADF) can be defined as:

$$BSADF_{r_2}(r_0) = \sup_{s_1 \in [0, r_2 - r_0]} ADF_{s_1,s_2} \qquad (2)$$

In other words, $ADF_{s_1,s_2}$ is calculated for the recursive window of $[s_1, s_2]$ until observation $s_2$. If the test statistic exceeds the critical value, then this observation belongs to a phase of explosive behavior. Compared to a single ADF test with window of $[0, s_2]$, PSY argue that this produced is more general and thus more effective at identifying bubbles when there can be multiple episodes during the sample period. Using daily cryptocurrency price data, we identify the date timestamped by the algorithm as bubble and visualize the outcome. This algorithm is also used by Corbet, Lucey and Yarovaya (2018) and Bouri, Shahzad and Roubaud (2019) to detect bubbles in various cryptocurrencies.

**2.2 Cryptocurrency-LAND Wealth Effect**

To calculate LAND price time series, we use the hedonic pricing model which is the workhorse of real estate price indices (Fisher, Geltner and Webb, 1994; Hill, 2013), which we refer to as the hedonic price index (HPI). The hedonic pricing model typically follows Equation 3, where log prices $p_i$ are regressed on relevant attributes $x_{ij}$ that could affect the transaction prices along with time fixed effects $\delta_t$, whose coefficients are used to construct the indices.

$$p_{it} = \sum_t \delta_t + \sum_j \beta_j x_{ij} + \varepsilon_{it} \tag{3}$$

In essence, the HPI represents changes in the conditional average of LAND prices in each period. To ensure that there is sufficient data to estimate the coefficients, we construct the index at the weekly level. For both Decentraland and The Sandbox, we include log number of plots and indicator variable for transactions settled in wETH as control variables. Once the HPIs are constructed for each metaverse, we investigate the wealth effect by analyzing the lead-lag relationship between the cryptocurrency price and its respective LAND price using the Granger (1969) causality test. The vector autoregression (VAR) of lag $p$ is written as:

$$y_t = c + A_1 y_{t-1} + \cdots + A_p y_{t-p} + e_t \tag{4}$$

To ensure stationarity, we first difference the variables and conduct the ADF test to verify. In the baseline specification, we include the differenced LAND price index and cryptocurrency price in the vector $y$. In the extended specification, we also include BTC and ETH prices to control for general market conditions.

### 3. Results

**3.1 Bubble Timestamping**

First, we demonstrate the existence of bubbles using the PSY algorithm in MANA and SAND. The daily time series of cryptocurrency prices and phases of explosive bubbles between January 1, 2021, and August 12, 2022 are displayed in Figure 1. The vertical line in each panel corresponds to the October 28, 2021, when Facebook announced its decision to change its name to Meta, underpinning its interest in metaverse. During this period, MANA and SAND prices increased by 28.6 and 287.5 times respectively, and much of the increase occurred after the announcement. The PSY algorithm identifies at least 10 of the days between the announcement and the end of November 2022 as bubble. The result is similar to Philippas et al. (2019) who document BTC price jumps following arrival of signals derived from Twitter and Google Trends. During this period, 5.6% and 5.3% of the days are identified as bubbles for MANA and SAND, while BTC is never identified as in a bubble, and ETH identified as bubble 3% of the days, albeit in different periods from MANA and SAND. This evidence shows that the MANA/SAND bubble is related to the interest in metaverse.

The result on bubble identification highlights an interesting aspect of the PSY algorithm. Shahzard, Anas and Bouri (2022) use a slightly different version of the algorithm and 4-hourly price data between January 1, 2020 to June 18, 2021 to investigate the impact of Elon Musk's tweets on Bitcoin and Dogecoin prices. While our analysis does not indicate a presence of bubble in BTC, their analysis identifies multiple episodes between January and April – often lasting several days – as bubbles. The inclusion of data from 2021 and analysis of higher frequency data will likely affect our results and more periods may be identified as bubbles, but the key insight from our analysis is that the metaverse bubble is not related to general market movements. Thus, early buyers of MANA and SAND are beneficiaries of this bubble.

**Figure 1: Cryptocurrency Bubble Timestamping**
This figure plots the daily prices of Decentraland's MANA, The Sandbox's SAND, Bitcoin (BTC), the Ethereum blockchain's Ether (ETH) between January 1, 2021 and August 12, 2022. The shaded regions are dates identified by the Phillips, Shi and Yu (2015) algorithm as explosive bubbles. On October 28, 2021, Facebook announced its decision to change the company's name to Meta, which is denoted by the vertical line in each plot. The proportion of days identified as in bubbles is reported in Table 1.

Panel A: Decentraland's MANA

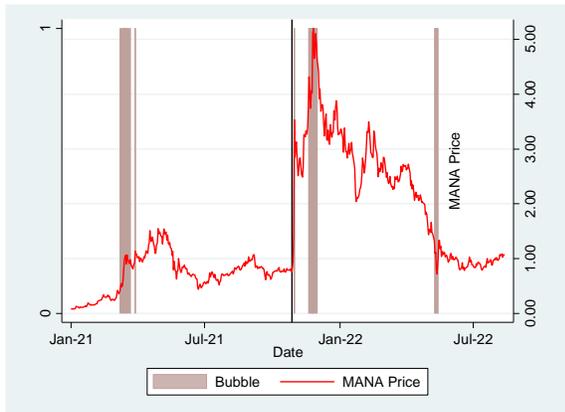

Panel B: The Sandbox's SAND

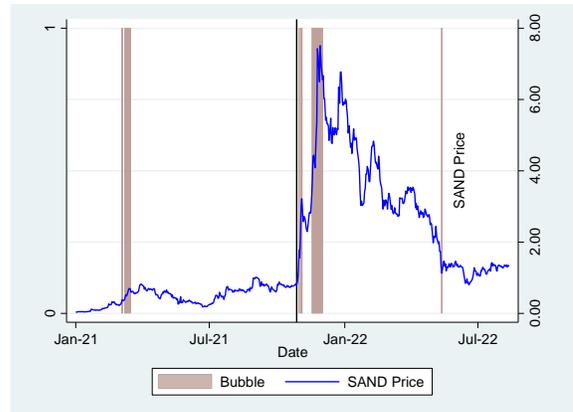

Panel C: Bitcoin (BTC)

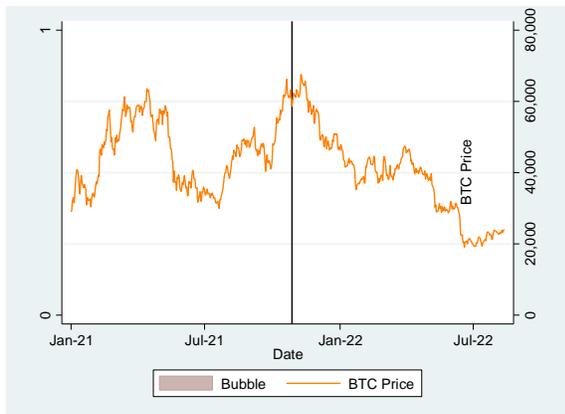

Panel D: Ether (ETH)

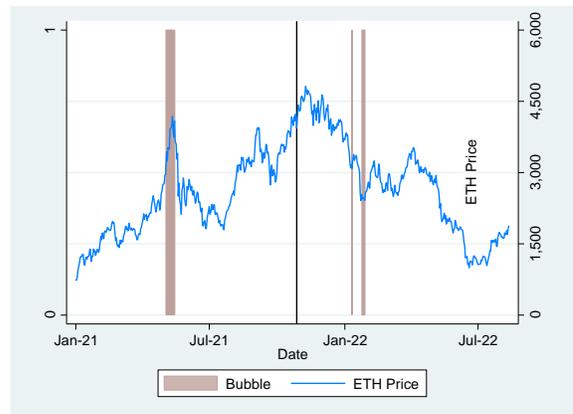

## 3.2 Cryptocurrency-LAND Wealth Effect

The explosive increase in metaverse cryptocurrencies coincide with the increase in LAND NFT prices. We estimate the HPI (base = 1) for LAND in Decentraland and The Sandbox and report the summary statistics of the HPI and corresponding cryptocurrencies at weekly frequency in Table 3. In figure 2, we plot the weekly HPI level (bar) against its cryptocurrency price (line). The plots show a very strong co-movement: weekly LAND-MANA correlation is 0.9870 and the LAND-SAND correlation is 0.9684. Both MANA and SAND are highly correlated (0.9675), and so are their HPI (0.9463). Because of this strong correlation, if the cryptocurrencies are considered to be in a bubble, then so can the metaverse real estate.

**Table 3: Summary Statistics for Price Indices**

Panel A reports the summary statistics of estimated hedonic price index (HPI) and its corresponding metaverse cryptocurrency at weekly frequency. Panel B reports the pairwise correlation between the HPI and cryptocurrencies. Decentraland's MANA was launched around the same time as its LAND NFTs, while The Sandbox's LAND NFTs were available for sale prior to the launch of SAND. Prior to the launch, The Sandbox accepted other cryptocurrencies and crypto assets as payments.

Panel A: Summary statistics

|          | Decentraland | | The Sandbox | |
| --- | --- | --- | --- | --- |
|          | HPI   | MANA | HPI    | SAND |
| N        | 206   | 206  | 140    | 104  |
| Mean     | 5.11  | 0.62 | 48.63  | 1.40 |
| Min      | 0.55  | 0.02 | 0.50   | 0.04 |
| Median   | 1.60  | 0.08 | 14.95  | 0.68 |
| Max      | 28.63 | 5.12 | 287.48 | 7.43 |
| Std Dev  | 6.58  | 0.98 | 76.90  | 1.70 |
| Skewness | 1.83  | 2.06 | 1.77   | 1.52 |
| Kurtosis | 5.35  | 6.89 | 4.73   | 4.55 |

Panel B: Pairwise correlation

|                 | DL HPI | MANA   | SB HPI | SAND   |
| --- | --- | --- | --- | --- |
| Decentraland HPI | 1.0000 |        |        |        |
| MANA            | 0.9780 | 1.0000 |        |        |
| The Sandbox HPI | 0.9506 | 0.9463 | 1.0000 |        |
| SAND            | 0.9309 | 0.9675 | 0.9674 | 1.0000 |

**Figure 2: Cryptocurrencies Prices and Hedonic Price Indices**
Weekly LAND HPI and the price of its corresponding metaverse cryptocurrency are plotted as bars and lines respectively. MANA and SAND are cryptocurrencies issued by the metaverse which can be used to purchase in-game LAND and other crypto assets. On October 28, 2021, Facebook announced its decision to change the company's name to Meta, which is denoted by the vertical line in each plot.

Panel A: Decentraland        Panel B: The Sandbox

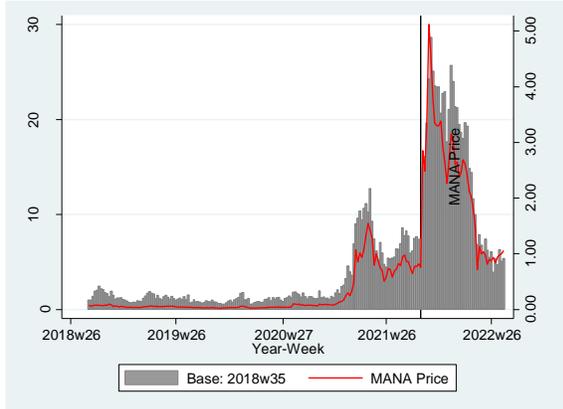
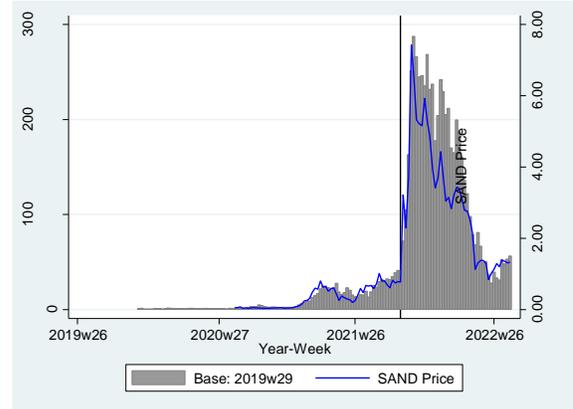

The wealth effect in our research design corresponds to cryptocurrency price leading LAND price. We begin by analyzing the lagged correlation between the price levels, a method often used in price-volume correlation and bubble research (e.g. DeFusco, Nathanson and Zwick, 2022; Liao, Peng and Zhu, 2022; Pénasse and Renneboog, 2022).[7] The idea is that correlation between current level of HPI and the lead (negative number) or lag (positive number) cryptocurrency prices can indicate the direction of the relationship. If cryptocurrency price leads LAND price, correlation should be higher for lag and decays more slowly as lag increases. Thus, the lefthand side should be lower and steeper. Figure 3 plots the correlations at weekly frequency for up to 10 weeks. The results indicate that cryptocurrency prices lead LAND prices for both Decentraland and The Sandbox, consistent with the wealth effect prediction.

---

[7] A comprehensive review of how behavioral biases can lead to the relationship between price and volume in different assets can be found Barberis (2018).

**Figure 3: Cryptocurrency-LAND Lead-Lag Correlation**
This figure plots the correlation between weekly HPI level and its corresponding cryptocurrency price at lag k. Positive numbers correspond to lags, while negative numbers correspond to leads. The correlation for Decentraland is plotted as solid line, while the correlation for The Sandbox is plotted as dashed line.

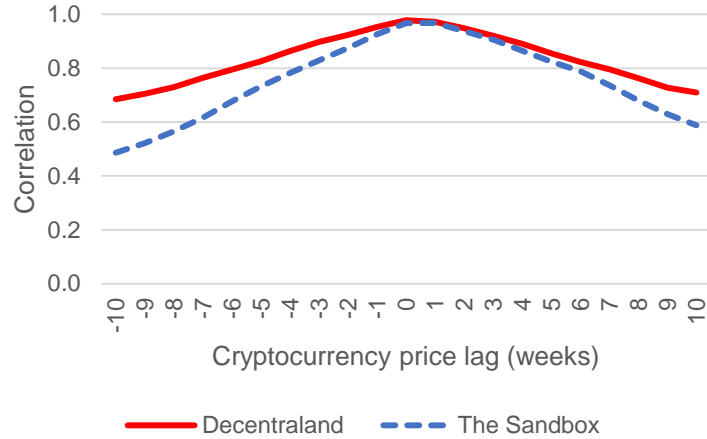

### 3.3 Granger Causality Test

The VAR Granger causality test allows a more rigorous detection of the lead-lag relationship between cryptocurrency and LAND prices. In Panel A of Table 4, we report the summary statistics of the first-differenced, weekly data. The ADF test statistics confirm the stationarity. The pairwise correlations of the first-differenced series are lower compared to level series, but the cryptocurrency prices are still highly correlated to LAND prices (0.4693 for Decentraland and 0.4172 for The Sandbox). LAND prices are also correlated to BTC and ETH, but to lesser extent.

Panel C and D show the Granger causality test for Decentraland and The Sandbox for up to 3 lags. In the extended VAR, we also include both BTC and ETH prices to control for market movements. In all lags and specifications, cryptocurrency prices lead LAND prices, where p-values for the cryptocurrency-to-LAND direction are below 1 percent level for most specifications, while the p-values for the reverse direction always exceed 10 percent, providing support for our wealth effect interpretation. Our result is similar to Goetzmann, Renneboog and Spaenjers (2011), who find that equity market returns have a strong, positive relationship with art market index between 1840 and 2007, and complements Dowling (2022b) with the direct investigation of co-movement between LAND NFTs and their corresponding cryptocurrencies.

**Table 4: Granger Causality Test and the Wealth Effect**
This table reports the various results leading to VAR Granger causality test. Panel A reports the summary statistics of the first-differenced, weekly time series along with the statistics of the augmented Dickey and Fuller (1979) test of unit root. Panel B reports the pairwise correlation of the first-differenced data. Panel C and D report the F-test statistic of the test that lagged value of cryptocurrencies Granger causes LAND prices at up to 3 lags for Decentraland and The Sandbox respectively. In the extended version, the time series of Bitcoin and Ether prices are also included.

Panel A: Summary statistics of first-differenced time series

|  | DL HPI | MANA | SB HPI | SAND | BTC | ETH |
|---|---|---|---|---|---|---|
| N | 204 | 204 | 138 | 102 | 204 | 204 |
| Mean | 0.8% | 1.3% | 2.9% | 3.1% | 0.6% | 0.9% |
| Min | -77.3% | -71.9% | -65.0% | -64.6% | -45.4% | -59.3% |
| Median | -0.2% | 0.4% | 2.2% | -0.9% | 0.4% | 0.4% |
| Max | 70.1% | 132.8% | 75.3% | 143.1% | 28.7% | 47.3% |
| Std Dev | 21.2% | 20.0% | 23.3% | 27.4% | 10.6% | 14.4% |
| Skewness | -0.14 | 1.67 | 0.14 | 1.23 | -0.64 | -0.71 |
| Kurtosis | 4.23 | 13.80 | 3.50 | 8.70 | 5.73 | 5.68 |
| ADF test statistic | -16.81 | -15.40 | -12.10 | -10.72 | -14.16 | -14.70 |
| p-value | 0.000 | 0.000 | 0.000 | 0.000 | 0.000 | 0.000 |

Panel B: Pairwise correlation

|  | DL HPI | MANA | SB HPI | SAND | BTC | ETH |
|---|---|---|---|---|---|---|
| Decentraland HPI | 1.0000 |  |  |  |  |  |
| MANA | 0.4693 | 1.0000 |  |  |  |  |
| The Sandbox HPI | 0.1931 | 0.2403 | 1.0000 |  |  |  |
| SAND | 0.5340 | 0.8113 | 0.4172 | 1.0000 |  |  |
| BTC | 0.3190 | 0.5227 | 0.2475 | 0.4546 | 1.0000 |  |
| ETH | 0.3012 | 0.5491 | 0.3104 | 0.4885 | 0.8461 | 1.0000 |

Panel C: Granger causality test for Decentraland

| Lag (weeks) | BTC & ETH | MANA → LAND F-stat | p-value | LAND → MANA F-stat | p-value | d.f. | Resid. d.f. | N |
|---|---|---|---|---|---|---|---|---|
| 1 | Not inc. | 15.96 | 0.000 | 2.25 | 0.135 | 1 | 201 | 204 |
| 2 | Not inc. | 12.86 | 0.000 | 0.96 | 0.384 | 2 | 198 | 203 |
| 3 | Not inc. | 8.54 | 0.000 | 1.61 | 0.187 | 3 | 195 | 202 |
| 1 | Included | 10.30 | 0.002 | 2.33 | 0.128 | 1 | 199 | 204 |
| 2 | Included | 10.17 | 0.000 | 0.73 | 0.485 | 2 | 194 | 203 |
| 3 | Included | 6.97 | 0.000 | 1.29 | 0.278 | 3 | 189 | 202 |

Panel D: Granger causality test for The Sandbox

| Lag (weeks) | BTC & ETH | SAND → LAND F-stat | p-value | LAND → SAND F-stat | p-value | d.f. | Resid. d.f. | N |
|---|---|---|---|---|---|---|---|---|
| 1 | Not inc. | 5.36 | 0.019 | 0.17 | 0.680 | 1 | 99 | 102 |
| 2 | Not inc. | 5.32 | 0.007 | 1.89 | 0.156 | 2 | 96 | 101 |
| 3 | Not inc. | 3.41 | 0.021 | 0.72 | 0.543 | 3 | 93 | 100 |
| 1 | Included | 6.15 | 0.015 | 0.58 | 0.448 | 1 | 97 | 102 |
| 2 | Included | 5.42 | 0.006 | 1.87 | 0.160 | 2 | 92 | 101 |
| 3 | Included | 3.19 | 0.028 | 0.57 | 0.639 | 3 | 87 | 100 |

Our result complements Nakavachara and Saengchote (2022), who find that transactions settled in SAND whose price increased more than other cryptocurrencies' are priced higher in USD. The authors interpret this as the effect of unit of account, but it could also be viewed as crypto wealth effect. Their analysis ended in January 2022, when the crypto market started declining from its peak in November 2021 before suffering a collapse in mid-May, so our extended sample until August 2022 offers a more complete view of the cycle. In the Appendix, we also investigate the relationship for the Otherside metaverse of Bored Ape Yacht Club (BAYC), the iconic NFT project and one of the three NFTs analyzed by Dowling (2022b). The project launched during the downward part of the cycle and exhibit the same lead-lag relationship, so the wealth effect works in both directions.

## 4. Conclusion

In this article, we document the wealth spillover effect from metaverse cryptocurrencies (MANA and SAND) into their corresponding virtual real estate (LAND NFTs). Real estate bubbles have occurred throughout history. In the case of these metaverses where opportunities to directly earn real estate income are not established, the situation reminisces the Florida real estate bubble in the mid-1920s. In an article discussing lessons from the 1920s crisis relevant to the 2008 crisis, White (2009) pointed to Galbraith's (1954) observation that rather than being "elements of substance", real estate prices were "based on the self-delusion that the Florida swamps would be wonderful residential real estate".

While Simpson (1933) surmised that the 1920s American real estate bubble was a result of a "dangerous" collaboration between banks, real estate promoters and local politicians, rather than wealth spillover effect from other sources, it nevertheless provided the necessary ingredient for leveraged positions in equity during the late 1920s. Thus, it could be argued that wealth spillover had its role in the 1929 crash. From the analysis in White (2009), the real estate-equity double bubbles spanned almost a decade, but in the crypto world where price movements are more volatile and change more rapidly, the wealth effect can take just weeks (if not days, as suggested by the BAYC's result in the Appendix) to occur. As evident from the 2008 crisis, evaporation of real estate wealth can cascade into a systemic crisis. For this metaverse cryptocurrency-LAND double bubble, the impact is likely limited as NFTs are not yet widely accepted as loan collateral, but our finding confirms that the fear outlined by the FSB is justified.[8]

Online communities often jest that the decentralized finance "experiment" is "speed running the evolution of the modern financial system".[9] but put differently, it also means we are reliving our mistakes from the past, and some are paying the price for it.

---

[8] One example of a lending platform which accepts NFTs as collateral is NFTfi, which offers loans denominated in wETH and DAI (a collateralized stablecoin). As of September 9, 2022, the majority of loans denominated in wETH are secured by Bored Ape Yacht Club and Mutant Ape Yacht Club NFTs, while loans denominated in DAI are mostly secured by wrapped Cryptopunks and Bored Ape Yacht Club. Loans collateralized by LAND NFTs account for a less than 1 percent of outstanding loans on the platform.

[9] See, for example, https://news.ycombinator.com/item?id=26262596, accessed on September 9, 2022.

**Appendix: Bored Ape Yacht Club's Otherside**

Bored Ape Yacht Club (BAYC) is arguably the most well-known digital art NFT with the highest price ever attained for a single sale is $2.85 million, albeit paid in ETH.[10] There are 10,000 unique BAYC NFTs, each with its own artistic attributes. The NFT contract was created on April 22, 2021. The ApeCoin (APE) cryptocurrency contract was created on February 14, 2022 and began trading in April, while the corresponding LAND NFT smart contract was launched slightly prior to the crypto market crash of May 2022 and LAND on May 1, covering only the downward part of the cycle.

We repeat the same analyses for Otherside, but since there are more transactions, the HPI can be estimated at daily frequency. Thus, we conduct the analyses at daily frequency and find the same results. The correlation between Otherside's HPI and APE is 0.9211, and the Granger causality test (with up to 5 lags because of more granular frequency) suggests that it takes at least 3 days before the spillover to occur, which provides comfort to the interpretation of wealth effect as consumers begin to feel the effect of the change in wealth and change their behavior. Overall, the BAYC's result suggests that the (negative) wealth effect is also present in the downward cycle.

**Table A1: Otherside LAND Transactions**

This table reports the summary statistics of secondary LAND transactions of Bored Ape Yacht Club's Oterside obtained from the Ethereum blockchain. Transactions occurred from May to August 2022. Prices are converted to USD using daily prices obtained from CoinGecko. Number of plots is the number of LAND NFTs that change ownership in a single transaction. Transactions settled in wETH (smart contract, "wrapped" version of Ether) are more likely to correspond to auction bids rather than fixed-price purchases and thus are explicitly flagged.

| Sample size | 16,110 | |
|---|---|---|
| Paid in wETH | 21.14% | |
| | USD price | Num plot |
| Mean | 14,324.77 | 1.04 |
| StdDev | 26,430.06 | 0.48 |
| Skewness | 6.28 | 17.98 |
| Kurtosis | 54.89 | 388.55 |
| p5 | 2,487.51 | 1 |
| p50 | 6,531.48 | 1 |
| p95 | 43,476.05 | 1 |

---

[10] https://cryptopotato.com/bored-ape-yacht-club-nft-sold-for-2-85-million-in-eth/, accessed on September 9, 2022.

## Figure A1: Cryptocurrency Bubble Timestamping

Panel A plots the daily prices of Bored Ape Yacht Club's ApeCoin (APE) between April 1 and August 12, 2022. The shaded regions are dates identified by the Phillips, Shi and Yu (2015) algorithm as explosive bubbles. Panel B plots the daily LAND HPI and APE price as bars and lines respectively. Because Otherside has greater daily activity compared to other metaverses, the HPI can be estimated at daily frequency. The correlation between HPI and APE is 0.9211.

Panel A: APE bubble timestamping

Panel B: LAND-APE correlation

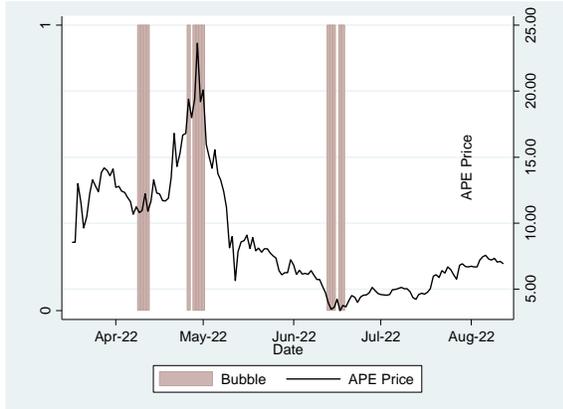
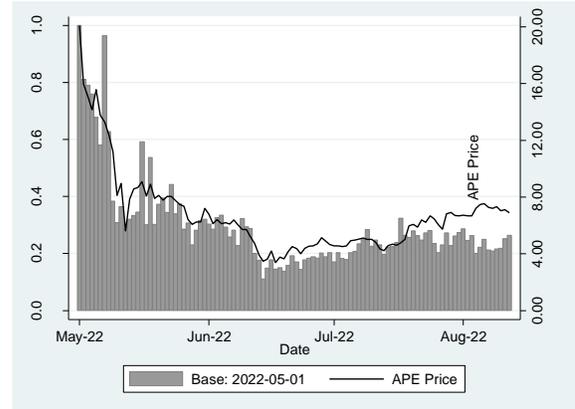

## Table A2: Granger Causality Test and the Wealth Effect

This table reports the various results leading to VAR Granger causality test. Panel A reports the summary statistics of the first-differenced, weekly time series along with the statistics of the augmented Dickey and Fuller (1979) test of unit root. Panel B reports the pairwise correlation of the first-differenced data. Panel C reports the F-test statistic of the test that lagged value of cryptocurrencies Granger causes LAND prices at up to 3 lags. In the extended version, the time series of Bitcoin and Ether prices are also included.

Panel A: Summary statistics of first-differenced data

|  | O HPI | APE | BTC | ETH |
|---|---|---|---|---|
| N | 103 | 103 | 103 | 103 |
| Mean | -1.3% | -1.0% | -0.4% | -0.4% |
| Min | -67.4% | -46.6% | -17.3% | -18.8% |
| Median | 2.3% | -0.5% | -0.2% | -0.6% |
| Max | 57.8% | 33.3% | 7.6% | 15.5% |
| Std Dev | 20.6% | 9.9% | 3.9% | 5.6% |
| Skewness | -0.22 | -0.87 | -0.91 | -0.10 |
| Kurtosis | 4.44 | 8.16 | 5.87 | 3.98 |
| Test Statistic | -15.82 | -12.53 | -10.10 | -10.05 |
| p-value | 0.0000 | 0.0000 | 0.0000 | 0.0000 |

Panel B: Pairwise correlation

|  | O HPI | APE | BTC | ETH |
|---|---|---|---|---|
| Otherside HPI | 1.0000 |  |  |  |
| APE | 0.4341 | 1.0000 |  |  |
| BTC | 0.4960 | 0.6712 | 1.0000 |  |
| ETH | 0.5030 | 0.6527 | 0.8990 | 1.0000 |

Panel C: Granger causality test for BAYC Otherside (daily frequency)

| Lag | BTC & ETH return | APE → LAND F-stat | p-value | LAND → APE F-stat | p-value | d.f. | Resid. d.f. | N |
|---|---|---|---|---|---|---|---|---|
| 1 | Not inc. | 1.35 | 0.25 | 0.21 | 0.645 | 1 | 100 | 102 |
| 2 | Not inc. | 2.25 | 0.11 | 0.74 | 0.480 | 2 | 97 | 101 |
| 3 | Not inc. | 4.12 | 0.01 | 0.58 | 0.632 | 3 | 94 | 100 |
| 4 | Not inc. | 4.11 | 0.00 | 0.54 | 0.710 | 4 | 91 | 99 |
| 5 | Not inc. | 4.05 | 0.00 | 0.78 | 0.564 | 5 | 88 | 98 |
| 1 | Included | 1.25 | 0.27 | 0.48 | 0.488 | 1 | 98 | 102 |
| 2 | Included | 1.78 | 0.17 | 1.22 | 0.301 | 2 | 93 | 101 |
| 3 | Included | 2.86 | 0.04 | 0.98 | 0.406 | 3 | 88 | 100 |
| 4 | Included | 3.27 | 0.02 | 0.81 | 0.524 | 4 | 83 | 99 |
| 5 | Included | 4.58 | 0.00 | 1.07 | 0.382 | 5 | 78 | 98 |